\documentclass[american,prd,preprint,a4paper,showpacs,amsfonts,amssymb]{revtex4}
\usepackage{amsmath}
\usepackage{babel}
\usepackage{bm}

\begin{document}

\newcommand\sla[1]{#1\hspace{-8pt}/\hspace{4pt}}
\newcommand\slas[1]{#1\hspace{-5pt}/}



\preprint{ZMP-HH/05-18}

\title{Covariant canonical quantization}

\author{Georg M. von Hippel}
\email{vonhippg@uregina.ca}
\affiliation{Department of Physics, University of Regina,
             3737 Wascana Parkway, Regina, Saskatchewan S4S 0A2, Canada}

\author{Mattias N.R. Wohlfarth}
\email{mattias.wohlfarth@desy.de}
\affiliation{{II.} Institut f\"ur Theoretische Physik, Universit\"at Hamburg,
             Luruper Chaussee 149, 22761 Hamburg, Germany} 

\begin{abstract}
We present a manifestly covariant quantization procedure based on the
de Donder--Weyl Hamiltonian formulation of classical field
theory. This procedure agrees with conventional
canonical quantization only if the parameter space is $d=1$
dimensional time. In $d>1$ quantization requires a fundamental length
scale, and any bosonic field generates a spinorial wave function,
leading to the purely quantum-theoretical emergence of spinors as a
byproduct. We provide a probabilistic interpretation of the wave
functions for the fields, and apply the formalism to a number of
simple examples. These show that covariant canonical quantization
produces both the Klein-Gordon and the Dirac equation, while
also predicting the existence of discrete towers of identically
charged fermions with different masses. Covariant canonical
quantization can thus be understood as a `first' or pre-quantization
within the framework of conventional QFT.
\end{abstract}
\maketitle


\section{Introduction}

The apparent incompatibility between general relativity and quantum
mechanics has long been a topic of concern and interest in the
theoretical physics community. Diffeomorphism invariance has to be
satisfied on the side of a general relativistic theory, in particular
denying any fundamental distinction between the notions of space and
time; but it is less clear how to achieve this requirement in, or
properly translate it to, a quantum theory. This particularly applies
to the canonical formulation of quantum mechanics and quantum field
theory based on a Hamiltonian treatment. A neat way around this
problem may be seen in path integral quantization which explains why
the predictions of the quantized theory still possess the relativistic
symmetries of the classical theory; but from the Hamiltonian point of
view with its explicit space-time split this is not a special merit of
the quantization procedure. This motivates the question whether there
is a covariant extension of Hamiltonian methods which also allows for
a manifestly covariant quantization procedure.
 
On the level of classical field theory there is indeed a Hamiltonian
formulation that does not rely on singling out a time coordinate, but
treats all spacetime coordinates equally throughout. This theory was
presented already in the nineteen-thirties by de
Donder~\cite{deDonder:1935} and Weyl~\cite{Weyl:1935}. Full covariance
is maintained through the use of multi-momenta, where one momentum is
associated to each partial derivative of the fields. While providing a
fully covariant equivalent to the standard Hamiltonian formulation of
field theory (in the sense of providing the same solutions), the de
Donder--Weyl formulation of classical dynamics has not received too
much attention. Only recently have there been several attempts to
quantize field theories on its basis. An early attempt by
Good~\cite{Good:1994,Good:1995} has been shown to disagree with
ordinary quantum mechanics and to give incorrect predictions for the
hydrogen spectrum~\cite{Navarro:1995sm}. Subsequently, a quantum
equation based on de Donder-Weyl theory has been conjectured by
Kanatchikov~\cite{Kanatchikov:1996fx} and
Navarro~\cite{Navarro:1998en}. There have also been attempts to
obtain a path integral formulation
\cite{Giachetta:2004yb,Bashkirov:2004kn,Bashkirov:2004bz} and a
version of Bohmian mechanics \cite{Nikolic:2004cv,Nikolic:2005nk}
based on de Donder--Weyl dynamics. Other recent applications of the de
Donder--Weyl formulation of field theory include a derivation of the
Ashtekar-Wheeler-DeWitt equation of canonical quantum
gravity~\cite{Rovelli:2002ef}.

In this paper, after a brief review of some of the elements of
the classical de Donder-Weyl theory in the following section
\ref{sec.classical}, we will formulate a covariant Poisson
bracket. In section \ref{sec.quantum} we will then proceed to
apply the Dirac quantization postulate to the latter. If supplemented
with a second, geometrically motivated, quantization postulate, this
leads to the same quantum evolution equation that had previously
been conjectured on the basis of analogies
\cite{Kanatchikov:1998xz,Navarro:1998en}.
Our approach for the first time presents a derivation of this
equation, which unifies both the Schr\"odinger and the Dirac equation,
from first principles. We go on to develop the quantum theory in the
covariant Schr\"odinger picture; in particular, we will discuss
the representation of operators, the consequences of an
indefinite scalar product on the Hilbert space which immediately
follows from the requirements of covariance, and the probability
interpretation of the wave functions. We apply the theory to a
number of basic problems in section \ref{sec.applications}, with
sometimes surprising results:

Among them are a new derivation of the Klein-Gordon equation that
makes no use of the relativistic energy-momentum relation, the
emergence of spinors from the quantization of scalar theories, and in
particular the emergence of the Dirac equation from the quantization
of any scalar field action. This means that the quantization procedure
here presented does not replace quantum field theory; instead, it is
found to provide a supplementary ``first quantization''. A result of
potential phenomenological interest is the prediction of towers of
identically charged fermions that differ only by their masses,
providing a qualitative explanation for the generations in the
Standard Model.


\section{Covariant Hamiltonians in classical field theory}\label{sec.classical}

This section reviews the covariant Hamiltonian treatment of
classical field theories, discussed first by de Donder and Weyl
\cite{deDonder:1935,Weyl:1935}, which is based on the introduction of
multi-momenta
associated to the partial derivatives of the fields. We then define
a new covariant Poisson bracket to rewrite the general phase space
evolution equations in an equivalent form suitable for quantization.

\subsection{Field theory in the multi-symplectic formalism}

Consider a geometrically well-defined field theory which is
diffeomorphism invariant on a $d$-dimensional Lorentzian background
manifold $\Sigma$. We will call this background manifold the parameter
space of the theory, coordinatized by parameters
$\left\{\sigma^a\right\}$ with corresponding partial derivatives
$\partial_a=\partial/\partial\sigma^a$.  Classical fields $q^i$ are
functions on this manifold, i.e.,  
\begin{equation}
q^i:\,\Sigma\rightarrow\mathbb{R}\,.
\end{equation}
In some theories it is convenient to consider a set of $n$ fields
$\left\{q^i\right\}$ as 
coordinates of a second, $n$-dimensional, target space manifold $M$;
in this case, requiring the theory to be geometrically well-defined
means it should obey the further diffeomorphism invariance on
$M$. The notion of a target space manifold is, however,
secondary. We define the theory on $\Sigma$ by its action,
which is obtained from the integration over $\Sigma$ of a scalar
Lagrangian as
\begin{equation}\label{eq.theory}
S=\int_\Sigma d^d\sigma \sqrt{-g}\,L\left(q^i,\nabla_a q^j\right).
\end{equation}
Note that the standard quadratic kinetic term in the Lagrangian
depends on $\nabla_a q^i$. Forming a scalar from these
(covariant) derivatives necessitates the existence of a
non-degenerate, and hence invertible, metric $g$ on $\Sigma$, the
signature of which we take to be~$(-,+,\dots,+)$. 
The determinant of this metric appears in the integration measure.
An explicit dependence of $L$ on the coordinates of $\Sigma$ is excluded 
by the requirement of diffeomorphism invariance.

The equations of motion of the theory (\ref{eq.theory}) are the
Euler-Lagrange equations derived by variation of the action with
  respect to the fields,
\begin{equation}\label{eq.euler}
\nabla_a\frac{\partial L}{\partial \nabla_a q^i}-\frac{\partial
  L}{\partial q^i}=0\,,
\end{equation}
where the covariant derivative involves the unique torsion free
and metric compatible Levi-Civita connection of $g$. The
necessary boundary condition requires a vanishing integral 
\begin{equation}
\int_{\partial \Sigma} dS^a \frac{\partial
  L}{\partial\nabla_aq^i}\delta q^i=0\,.
\end{equation}

The above equations of motion are partial differential equations of
second order, for first order Lagrangians. To reduce the order, the
standard Hamiltonian treatment introduces canonical momenta
${p_i=\partial L/\partial\partial_0 q^i}$. Clearly, these momenta
are non-covariant 
quantities as their definition explicitly depends on the choice of
time and hence on the choice of coordinate system on $\Sigma$. It
follows that the usual Hamiltonian function, depending on the
non-covariant canonical momenta, cannot be a scalar.

To remedy this apparent difficulty, we introduce the manifestly
covariant multi-momenta associated to each partial derivative of the
fields,
\begin{equation}
p^a_i=\frac{\partial L}{\partial\nabla_a q^i}\,, 
\end{equation}  
which transform as the components of a vector in the parameter
space tangent bundle $T\Sigma$ (and as those of a
differential form in $T^* M$, if the fields form coordinates of a
target space manifold). We assume Lagrangians such that the
multi-momenta as functions of the fields and their partial
derivatives may be solved for these derivatives to yield $\nabla_a
q^i\left(q^j,p^b_k\right)$. In terms of the new covariant momenta, we
may then also define the covariant Hamiltonian
\begin{equation}\label{ham}
H= p^a_i\nabla_a q^i-L 
\end{equation}
which is a function of the new independent variables $q^i$ and
$p^a_i$, and transforms as a diffeomorphism scalar on the
parameter space $\Sigma$.

The Euler-Lagrange equations imply the covariant Hamiltonian equations
\begin{subequations}\label{covHeqm}\begin{eqnarray}
\frac{\partial H}{\partial q^i} & = & -\nabla_a p^a_i\,,\\
\frac{\partial H}{\partial p^a_i} & = & \nabla_a q^i\,. 
\end{eqnarray}\end{subequations}
Conversely, given a covariant Hamiltonian $H\left(q^i,p^a_j\right)$,
we may define a Lagrangian $L\left(q^i,\nabla_a q^j\right)$ via
(\ref{ham}). Then the covariant Hamiltonian equations imply the
Euler-Lagrange equations. Diffeomorphism invariance again implies that
the Hamiltonian cannot depend explicitly on the coordinates of
$\Sigma$.  Below we will see that the covariant Hamiltonian
formalism nicely reduces to conventional Hamiltonian mechanics
if the parameter space $\Sigma$ is one-dimensional. 

\subsection{The classical Dirac field as an example}\label{DiracExample}
As an example for the powerful finite-dimensional phase space
formalism of de Donder and Weyl, we take a brief look at the massive
Dirac field. The Lagrangian in its symmetrical form is given by
\begin{equation}
L=\frac{1}{2}\bar{\psi}\gamma^a\nabla_a\psi-\frac{1}{2}\nabla_a\bar{\psi}\gamma^a\psi-M\bar\psi\psi\,,
\end{equation}
where we have introduced the Dirac matrices $\gamma^a$ of the (curved)
background, on which we will comment in more detail below. We treat
$\psi$ and $\bar\psi$ as independent, such that the conjugate covariant
momenta follow by definition as
\begin{equation}\label{DiracConstraints}
\pi^a_\psi=\frac{1}{2}\bar{\psi}\gamma^a\,,\qquad\pi^a_{\bar\psi}=-\frac{1}{2}\gamma^a\psi\,.
\end{equation}
These relations are, in fact, primary constraints, 
relating the spinors and their conjugate momenta. Although 
these momenta are not invertible to obtain 
$\nabla_a\psi(\psi,\bar\psi,\pi^b_\psi,\pi^c_{\bar\psi})$, and
similarly $\nabla_a\bar\psi$, we can define the covariant Hamiltonian
as
\begin{equation}\label{DiracHamiltonian}
H = M\bar{\psi}\psi +\left(\pi^a_\psi-\bar\psi\gamma^a/2\right)\lambda_a+\bar\lambda_a\left(\pi^a_{\bar\psi}+\gamma^a\psi/2\right),
\end{equation}
where the constraints have been added with the help of spinorial
Lagrange multipliers~$\lambda_a$ and $\bar\lambda_a$. The Dirac
equation and its conjugate now follow immediately from the covariant
Hamiltonian equations (\ref{covHeqm}) above, utilizing the constraint
equations.

\subsection{Definition of a covariant Poisson bracket}
With the aim of facilitating an easier transition to a quantum theory,
we consider Poisson brackets in the new formalism. A covariant
extension of the standard Poisson bracket is given by the definition
\begin{equation}
\left\{f,g\right\}_a=\frac{\partial f}{\partial
  q^i}\frac{\partial g}{\partial p^a_i}-\frac{\partial f}{\partial
  p^a_i}\frac{\partial g}{\partial q^i} 
\end{equation}
for any two phase space functions $f$ and $g$
depending on the Hamiltonian variables $q^i$ and $p^a_i$. This bracket
carries a further index, thus mapping two
functions of the canonical variables to a differential form in
$T^*\Sigma$. In general, it changes the number of indices and with it
the tensor structure defined by its arguments. This obstructs the
usefulness of this bracket definition, as valuable properties of the 
Poisson bracket are lost. This applies in particular to the
important Jacobi identity, which provides the algebra of phase
space functions with the structure
of a Lie algebra. Here the Jacobi identity is valid only for equal
subscripts, i.e., for 
expressions of the form $\{\{f,g\}_a,h\}_a$, but these are not allowed
as tensors on $\Sigma$.

Hence we are led to amending the bracket definition, and consider
brackets of the form
\begin{equation}
\left\{f,g\right\}= \left\{f,g\right\}_a t^a
\end{equation}
where we introduce an arbitrary vector field in $T\Sigma$
with components $t^a$, the origin of which we will discuss in the
following section on quantization. The only classical requirement that
we will make on this field concerns its normalization
$N(d)=g_{ab}t^at^b$ which may depend on the dimension of the parameter
spacetime $\Sigma$. It should be such that $N(1)=-1$ so that the usual
Poisson bracket may emerge when $d=1$ with $g_{\sigma\sigma}=-1$. 
It turns out that the bracket so defined satisfies the formal algebraic
properties of the Poisson bracket, which we state for phase space
functions $f$ and $g$ and real numbers $c$ (for other definitions of 
Poisson brackets within the de Donder--Weyl formalism, compare 
\cite{Forger:2002ak,Paufler:2002-49,Kanatchikov:1997wp}):

The Poisson bracket is antisymmetric and annihilates constants,\begin{subequations}
\begin{eqnarray}
\left\{f,g\right\} & = & -\left\{g,f\right\},\\
\left\{f,c\right\} & = & 0\,;
\end{eqnarray}
\end{subequations}
it is $\mathbb{R}$-linear in $f$ and $g$ (where linearity in the second
argument follows by antisymmetry),\begin{subequations}
\begin{eqnarray}
\left\{f_1+f_2,g\right\} & = &
\left\{f_1,g\right\}+\left\{f_2,g\right\},\\
\left\{cf,g\right\} & = & c\left\{f,g\right\};
\end{eqnarray}
\end{subequations}
the Poisson bracket further satisfies a product rule and, importantly, the Jacobi identity:\begin{subequations}
\begin{eqnarray}
\left\{f_1f_2,g\right\}  &=& 
\left\{f_1,g\right\}f_2\,+\,f_1\left\{f_2,g\right\},\\
\left\{\left\{f,g\right\},h\right\} &+&
\left\{\left\{g,h\right\},f\right\} \,+\,
\left\{\left\{h,f\right\},g\right\} \,=\,0\,.
\end{eqnarray}
\end{subequations}

The introduction of the vector field $t^a$ provides another advantage,
again with a view towards quantization: it allows us to achieve a one
to one correspondence between the fields~$q^i$ and the
contracted multi-momenta $p_i=-t_a p^a_i$. We find the covariant Poisson
brackets\begin{subequations}\label{classbra}
\begin{eqnarray}
\left\{q^i,q^j\right\} & = & 0\,,\\
\left\{p_i,p_j\right\} & = & 0\,,\\
\left\{q^i,p_j\right\} & = & -N(d)\delta^i_j\,.
\end{eqnarray}
\end{subequations}

Brackets including the covariant Hamiltonian generate the following
expressions, similar to those appearing in the covariant
Hamiltonian equations of motion (\ref{covHeqm}); the fact that there
is no precise agreement is due to the appearance of the vector field
$t^a$:
\begin{subequations}\label{eq.brackets}
\begin{eqnarray}
\left\{q^i,H\right\} & = & t^a\nabla_a q^i\,,\\
\left\{p_i,H\right\} & = & -N(d) \nabla_a p^a_i\,.\label{supp} 
\end{eqnarray}
\end{subequations}
These brackets are useful in evaluating the evolution equation for
phase space functions with respect to the parameters given by the
coordinates of $\Sigma$. We calculate
\begin{equation}
\left\{f,H\right\}  =  \frac{\partial f}{\partial q^i}t^a\nabla_a
q^i + \frac{\partial f}{\partial p^a_i}t^a\nabla_b p^b_i\,,
\end{equation}  
which may be rewritten as
\begin{equation}\label{evocl}  
t^a\nabla_a f - \left\{f,H\right\}  -t^a\nabla_a^0 f= \frac{\partial f}{\partial p^b_i}
\left( t^a\nabla_a
p^b_i - \left\{p^b_i,H\right\} \right). 
\end{equation}
This is the form of the general evolution equation which we will use
as an important ingredient of the quantization procedure. Note that the
parameter space derivatives of the phase space function~$f$ are
evaluated along the integral curves of the vector field
$t^a$. The derivative operator $\nabla_a^0$ acts only on the
$\sigma^a$-dependence of $f$ not coming in through the coordinates
and momenta. A closer inspection of the equation also reveals
that it is trivially satisfied for any phase space function
linear in the momenta, e.g., for $p^a_i$ or $t_ap^a_i$. This means we
have to supplement it with equation (\ref{supp}).

\subsection{Hamiltonian mechanics on one-dimensional $\Sigma$}
The results and constructions above are in complete analogy to the
standard Hamiltonian treatment of classical mechanics which is,
however, restricted to a one-dimensional parameter space $\Sigma$,
with time coordinate $\sigma$, if diffeomorphism invariance is
required.

The Hamiltonian formalism of de Donder and Weyl reduces to the standard one
for $d=1$. To see this more explicitly, note that the $T\Sigma$ index
$a$ of the multi-momentum $p^a_i$ can merely take a single value in
this case corresponding to the single coordinate $\sigma$ on $\Sigma$,
which may be suppressed. The manifold, its tangent and cotangent
spaces are all locally isomorphic to the real numbers. The
normalization requirement for the single-component vector field
enforces $t^\sigma=1$ because of our signature convention
$g_{\sigma\sigma}=-1$. Thus we
obtain agreement between our covariant Poisson bracket and the
standard one. Equations~(\ref{classbra}) reduce to the canonical
Poisson brackets, and equations (\ref{eq.brackets}) become equivalent
to the Hamiltonian equations. The right hand side of the phase space
evolution equation (\ref{evocl}) cancels; what remains is the
well-known time evolution formula 
\begin{equation}
\frac{d}{d\sigma}f-\left\{f,H\right\}-\frac{\partial}{\partial\sigma}f=0\,.
\end{equation}


\section{Covariant canonical quantization}\label{sec.quantum}

Following Dirac, the quantization of a system in classical mechanics
takes as its starting point the Hamiltonian formulation. The canonical
variables are promoted to operators acting on a Hilbert space, 
and the Poisson brackets to commutators. With our new covariant 
Hamiltonian formalism, we will now mimic these steps. 

\subsection{Quantization postulates}

As the bracket $\{f,g\}$ that we have defined above has the same
algebraic properties as the Poisson bracket, we promote it to a
commutator of operators in exactly the same way,
\begin{equation}\label{qupostul1}
\left\{f,g\right\}\quad\mapsto\quad -i l^{d-1}[\hat f,\hat g]\,. 
\end{equation}
According to a famous argument of Dirac \cite{Dirac}, this is in fact the
only consistent 
quantization postulate, if the quantum bracket is required to
preserve its classical algebraic properties. Phase space
functions $f$ and $g$ have been replaced by operators on some Hilbert
space, denoted by a hat. The imaginary unit is required to imply that
$-i\,[\cdot,\cdot]$ is self-adjoint for self-adjoint entries (with
respect to the Hilbert space inner product which we will define
below).
In our units, where $c=1$ and $\hbar=1$, we have to introduce another
independent length scale $l$, which we might choose to be the Planck
length $l_P$, to compensate the dimensions of the derivatives with
respect to the canonical variables that appear in our Poisson
bracket. This result does not depend on the dimension of the
fields $q^i$; it merely assumes the dimension of the Lagrangian
$L$ is $(\mathrm{length})^{-d}$. Note that the necessity 
of a fundamental length scale for quantization appears only on 
manifolds $\Sigma$ of dimension $d>1$. 

Now we have to think about the vector field $t^a$ in our bracket
definition. The obvious choice seems to be a classical
timelike vector field on $\Sigma$, with normalization ${N(d)=-1}$ for any
dimension. However, this would have several undesirable consequences; firstly,
no such vector field was included in the classical theory in the
original formulation (\ref{eq.theory}). Thus additional input would be
necessary for the quantum theory. Such input would not be universal in
the sense that the chosen timelike field could differ for different quantum
systems under consideration. Secondly, this would amount to introducing a
space-time split of $\Sigma$ into the product of a family of timelike
curves and their corresponding normal surfaces, thereby introducing
all of the problems associated with the canonical procedure.
Thus the aim of an intrinsically higher-dimensional quantization procedure
on $\Sigma$ would be lost. But what options are left now of choosing a
vector field which is implicitly given on any $\Sigma$?

To answer this question, we have to resort to some more geometry. On
every curved background manifold $\Sigma$ (admitting a spin structure)
exists an algebra of Dirac matrices $\gamma_a$ with the property that
\begin{equation}
\gamma_a\gamma_b+\gamma_b\gamma_a=2g_{ab}\,.
\end{equation}
As usual, these Dirac matrices are related to those of the local
Lorentzian tangent spaces~$\Gamma_\mu$ by the vielbeins $e_a^\mu$ as
$\gamma_a = e_a^ \mu \Gamma_ \mu$. The vielbeins form the metric as
$g_{ab}=e_a^\mu e_b^\nu\eta_{\mu\nu}$. The normalization of the Dirac
matrices gives $\gamma_a \gamma^a=d$. This leads us to propose the
following quantization postulate for the vector field $t^a$,
\begin{equation}\label{qupostul2}
t^a\quad\mapsto\quad -i n(d)\gamma^a\,,
\end{equation}
which implies $N(d)=-d n(d)^2$. To satisfy the requirements $N(1)=-1$
and $t^\sigma=1$ for the one-dimensional case, we note that the only
Dirac matrix in $d=1$ is $\Gamma^\sigma=i$, so that the normalization
function $n(d)$ must be chosen such that $n(1)=1$. Otherwise $n(d)$ is
quite arbitrary and must be fixed by application of the theory, which
remark also applies to the fundamental length scale $l$.

It is worth noting that the appearance of the Dirac matrices in this
context has a historical parallel in Dirac's original derivation of
the Dirac equation \cite{Dirac:1928hu,Dirac:1936tg}, where a universal
object with a covariant vector index was also required to fulfill the
demands of covariance.

\subsection{Quantum evolution -- Dirac is Schr\"odinger}\label{Derivation}

We will now motivate a quantum evolution equation based on our
classical covariant Hamiltonian picture, which turns out to unify
the Dirac and the Schr\"odinger equation.

The multi-symplectic phase space is spanned by the canonical
variables $q^i$ and $p^a_i$. Any `proper' phase space function $f$
depends on the coordinates of the parameter space $\Sigma$ only
through these variables. For the operators $\hat f$ associated
to such phase space functions we now analyze the requirement that
the classical evolution equation (\ref{evocl}) holds in its quantum
version as 
\begin{equation}
i l^{d-1}\,[\hat f,\hat H] - i\, n(d)\left(\sla{\nabla} \hat
f\right) = \mathcal{S}\left( \widehat{\frac{\partial
    f}{\partial p^a_i}}\left( i l^{d-1}[\hat p^a_i,\hat H] - i\, n(d) 
  \left(\sla{\nabla} \hat
p^a_i\right) \right)\right). 
\end{equation}
The operator $\mathcal{S}$ denotes a
symmetrization of the canonical variables, which are now operators,
and we use Feynman's shorthand notation $\sla{\nabla}=\gamma^a
\nabla_a$. Noting that $(\sla{\nabla}\hat f)=[\sla\nabla,\hat f]$ in
the action on states, the above equation is can be rewritten in the
form 
\begin{equation}
[\hat f,\hat H + n(d)l^{-d+1}\sla\nabla] = \mathcal{S} \left( 
\widehat{\frac{\partial f}{\partial p^a_i}} [\hat p^a_i,\hat
  H + n(d)l^{-d+1}\sla\nabla]\right). 
\end{equation}
This equation holds for all operators if, and only if, $\hat
H+n(d)l^{-d+1}\sla\nabla$ is a constant independent of the
canonical variables. For $d=1$, this reduces to $\hat
H+i\partial_\sigma$. To obtain the same result as in
conventional quantum mechanics in this limit, we have to set
$\hat H+n(d)l^{-d+1}\sla\nabla=0$. However, note that this type
of derivation in quantum mechanics does not produce the
Schr\"odinger equation, which is $\hat H-i\partial_\sigma=0$, acting
on Schr\"odinger picture states. This is because the quantization of
the classical evolution equation leads to an operator equation valid
in the Heisenberg picture. The unitary change of pictures is
responsible for the change of signs.

We assume, as will be justified in the following
section~\ref{Heisenberg}, as it requires some development of the
theory, that the same
change of signs occurs here. Thus we finally arrive at the quantum
evolution equation, as an equation acting on $\Sigma$-dependent
states: 
\begin{equation}\label{quevol}
\left(\hat H-n(d)l^{-d+1}\sla\nabla\right)\left|\psi(\sigma^a)\right>=0\,.
\end{equation}
Note that the quantum evolution effectively does not
require the terms appearing in the symmetrization operator, which
thus need not be specified. This is an advantage because it is not
consistently possible to do so even in conventional quantum
mechanics: no map of phase space functions into an operator algebra
exists, compatible with the Poisson bracket. 
An attempt to rectify this situation is made by deformation
quantization, employing as operators formal power series in the Planck
quantum $\hbar$, see \cite{Sternheimer:1998yg,Dito:2002dr} for
reviews.

The above equation gives the operator $\hat H$ the dimension of mass
times $l^{-d+1}$. Supposing, in an expansion in terms of the canonical
variables, that there is a constant term in $\hat H$, we find that
both the Schr\"odinger and the Dirac equation follow from the same
quantization procedure. The Schr\"odinger equation is relevant for a
quantization of fields
$x^i\,:\,\Sigma\cong\mathbb{R}\rightarrow\mathbb{R}^{n}$ and the Dirac
equation corresponds to quantized fields on a parameter
spacetime $\Sigma\cong M_{1,3}$. This will be further illustrated
below. Here we only note that, in $d>1$, the wave functions will
automatically become spinors.

The quantum evolution equation in the form (\ref{quevol}) has been
conjectured before by Kanatchikov \cite{Kanatchikov:1996fx} and
Navarro \cite{Navarro:2001ge} on the basis of analogies between the
Dirac equation and conventional quantum mechanics. Here we have
presented for the first time a physically reasonable derivation of
equation (\ref{quevol}) from first principles, based solely on
the two quantization postulates (\ref{qupostul1}) and
(\ref{qupostul2}).

\subsection{The local evolution operator and the Heisenberg picture}\label{Heisenberg}

We shall now introduce the covariant Heisenberg picture, and in
this context we will justify the remaining assumptions going into
the derivation of the quantum evolution equation~(\ref{quevol}). To
clarify the calculations we will use subscripts ${}_S$ and ${}_H$
for Schr\"odinger and Heisenberg picture quantities, respectively.

The first notion we need is that of an evolution operator $\hat
U(\sigma,\sigma_0)$, with the help of which a $\Sigma$-dependent
Schr\"odinger picture state may be written in terms of a
$\Sigma$-independent Heisenberg picture state: 
\begin{equation}
\left|\psi(\sigma)\right>_S=\hat U(\sigma,\sigma_0)\left|\psi(\sigma_0)\right>_H\,.
\end{equation}
The consistency of expectation values requires $U$ to be unitary in
the sense $\hat U^{\#}\hat U=\openone$ (cf. section \ref{Hilbert} below). 

On curved parameter spaces $\Sigma$, such an evolution operator can
only be defined locally, i.e., in a sufficiently small neighborhood of
a point $p\in\Sigma$. If $q$ is another point in this neighborhood,
then there is a unique geodesic joining $p$ and $q$. The corresponding
evolution operator $\hat U(q,p)$ can be written as $\hat
U(\sigma,\sigma_0)$ in a geodesic normal coordinate
system. Substituting the above state expansion into the quantum
evolution equation we find 
\begin{equation}
\hat H_S \hat U(\sigma,\sigma_0)-n(d)l^{-d+1}\sla{\nabla}\hat U(\sigma,\sigma_0)=0\,.
\end{equation} 
Solving this equation to first order in an infinitesimally
small displacement $\delta\sigma$ then gives
\begin{equation}\label{U1}
\hat U(\sigma_0+\delta\sigma,\sigma_0)=\openone+\frac{l^{d-1}}{dn(d)}\delta\sigma^a\gamma_a\hat H_S\,.
\end{equation}
The adjoint $\hat U^{\#}$, again to first order, follows from
$(\gamma_a\hat H_S)^{\#}=-\hat H_S\gamma_a$. Hence $\hat U$ is
unitary, if, and only, if 
\begin{equation}
[\gamma^a,\hat H_S]=0\,,
\end{equation}
which we must require. (The examples below show that this relation
usually holds.) The evolution operator for finite coordinate
differences within the local neighborhood follows from the limiting
procedure $\hat U(\sigma,\sigma_0)=\lim\, (\hat
U(\sigma,\sigma-\delta\sigma)\dots\hat
U(\sigma_0+\delta\sigma,\sigma_0))$ for $\delta\sigma\rightarrow 0$;
in consequence, it is unitary as well.

We are now in the position to calculate the equation of motion for
Heisenberg picture operators $\hat f_H =\hat U^{\#}\hat f_S \hat
U$. Using the same steps as in conventional quantum mechanics yields 
\begin{equation}
\sla{\nabla}\hat f_H-n(d)^{-1}l^{d-1}[\hat f_H,\hat H_H]=[\nabla_a\hat U^{\#},\gamma^a]\hat U\hat f_H\,.
\end{equation}
The right hand side of this equation does not contribute. This is
easily seen by noting that ${[\nabla_a\hat
    U^{\#},\gamma^a]=[\nabla_a\hat U,\gamma^a]^{\#}}$, and using
equation~(\ref{U1}) to determine ${\nabla_a\hat U\sim \gamma_a \hat
  H_S}$. Thus one finds $[\nabla_a\hat U,\gamma^a]\sim [\gamma_a,\hat
  H_S]$, the vanishing of which was required by the existence of the
Heisenberg-Picture. The Heisenberg equation of motion 
\begin{equation}\label{heisenevol}
n(d)l^{-d+1}\sla{\nabla}\hat f_H-[\hat f_H,\hat H_H]=0
\end{equation}
follows. Using the same method as in the derivation of the quantum
evolution equation in section~\ref{Derivation} one thus finds $\hat
H_H+n(d)l^{-d+1}\sla{\nabla}=0$ in the Heisenberg picture, fully
justifying the change of sign we made in obtaining
equation~(\ref{quevol}).

The Heisenberg equation of motion (\ref{heisenevol}) will play an 
important role in showing that our  theory indeed has the correct
classical limit in section \ref{Ehrenfest} below.

\subsection{Canonical operators in the Schr\"odinger picture}

In the covariant Schr\"odinger picture, all operators,
including the covariant Hamiltonian, act on states which are 
elements of some Hilbert space $\mathcal{H}$ and depend on the
coordinates of $\Sigma$, i.e. $\left|\psi(\sigma^a)\right>$.
We wish to find a realization of these operators acting on
wave functions in an explicit Schr\"odinger representation; for
this purpose we have 
to introduce a basis of $\mathcal H$, which is conveniently given by
states
\begin{equation}
\left|\bm q_\alpha\right>=\left|\bm q\right>\otimes e_\alpha\,,
\end{equation}
where $\left|\bm q\right>$ are eigenstates of the 
field operators $\hat q^i$,
i.e. $\hat{q}^i\left|\bm q\right> = q^i\left|\bm q\right>$, and $e_\alpha$ are
the canonical basis vectors of the representation space of the Dirac
algebra. The dual states are given by
\begin{equation}
\left<^\alpha\bm q\right|=\left<\bm q\right|\otimes \omega^\alpha\,,
\end{equation}
where $\{\omega^\alpha\}$ is the dual basis of $\{e_\alpha\}$, such
that the normalization condition becomes
\begin{equation}
\left<^\alpha\bm q\,|\,\tilde{\bm q}_\beta\right> =
\delta^\alpha_\beta\delta(\bm q-\tilde{\bm q})\,.
\end{equation}

In the basis $\{\left|\bm q_\alpha\right>\}$ any state of the Hilbert
space can be expanded as
\begin{equation}
\left|\psi\right> = \int d\bm q \left|\bm q_\alpha\right>
\psi^\alpha(\bm q)\,.  
\end{equation}
The components $\psi^\alpha(\bm q)=\left<^\alpha\bm q\,|\,\psi\right>$
with respect to this basis give the spinorial Schr\"odinger picture
wave function. In this notation we have suppressed the
$\Sigma$-dependence; more precisely, one should write $\psi^
\alpha(\sigma^a;\bm q)$. The identity operator on $\mathcal H$ has a
partition of the form
\begin{equation}
\openone = \int d\bm q \left|\bm q_\alpha\right>\left<^\alpha\bm q\right|\,.
\end{equation}

The canonical operators should satisfy the commutation relations
\begin{subequations}
\begin{eqnarray}
[\hat q^i,\hat q^j] & = & 0\,,\\
{[}\hat p_i,\hat p_j] & = & 0\,,\label{pprel}\\ 
{[}\hat q^i,\hat p_j] & = & idn(d)^2 l^{-d+1}
\delta^i_j\,,\label{qprel}
\end{eqnarray}
\end{subequations}
which follow from an application of the two quantization postulates to the
classical Poisson bracket equations (\ref{classbra}). It would
seem to be convenient at this stage to remove the dimension
dependence of the canonical commutation relations by setting
$n(d)=1/\sqrt{d}$, but we emphasize again that $n(d)$ should be
fixed by application. To study the action of the
canonical operators on wave functions, we need the following
matrix elements:
\begin{subequations}
\begin{eqnarray}
\left<^\alpha\bm q\,|\,\hat p_i\,|\,\tilde{\bm q}_\beta\right> &
= & idn(d)^2 l^{-d+1}\delta^\alpha_\beta \frac{\partial}{\partial \tilde
  q^i}\delta(\bm q-\tilde{\bm q}),\\
\left<^\alpha\bm q\,|\,\gamma^a\,|\,\tilde{\bm q}_\beta\right> & = &
(\gamma^a)^\alpha{}_\beta \delta(\bm q-\tilde{\bm q})\,,
\end{eqnarray}
\end{subequations}
where the first identity follows from an expansion of
$\left<^\alpha\bm q\,|\left[\hat q^i,\hat p_i\right]|\,\tilde{\bm
  q}_\beta\right> $, using (\ref{qprel}). Now we act with our
operators on arbitrary states, which yields
\begin{subequations}
\begin{eqnarray}
\hat{\bm q}\left|\psi\right> & = & \int d\bm q \left|\bm
  q_\alpha\right> \left(\bm q \psi^\alpha(\bm q)\right),\\
\hat{\bm p}\left|\psi\right> & = & \int d\bm q \left|\bm
  q_\alpha\right> \left(-idn(d)^2 l^{-d+1}\frac{\partial}{\partial \bm
  q}\psi^\alpha(\bm q)\right),\\ 
\gamma^a\left|\psi\right> & = & \int d\bm q \left|\bm
  q_\alpha\right> \left((\gamma^a)^\alpha{}_\beta \psi^\beta(\bm q)\right).
\end{eqnarray}
\end{subequations}
Thus $\hat{q}^i$ and $\gamma^a$ act multiplicatively on
wave functions, whereas the $\hat p_i$ essentially act as derivative
operators. The commutation relations stated above are clearly
satisfied

A further important point is missing for a successful
transition from the classical to the quantum theory. The classical
covariant Hamiltonian depends on the phase space variables~$q^i$ 
and~$p^a_i$, so that the Hamiltonian operator would seem to depend on
the operators $\hat p^a_i$ for which we have not yet given a
representation. However, it is always possible to replace these
operators by $\hat p_i$, as we will now show. Acting on
wave functions the $\hat p^a_i$ are realized by
\begin{equation}
\hat p^a_i \sim
- n(d)l^{-d+1}\gamma^a\frac{\partial}{\partial q^i}\,,
\end{equation} 
which may be derived from an application of our quantization
postulates to the classical Poisson bracket
$\{q^i,p^a_j\}=t^a\delta^i_j$. It follows that any occurrence of
$\hat{p}^a_i$ can be replaced by
\begin{equation}
\hat{p}^a_i= - \frac{i}{dn(d)}\gamma^a \hat{p}_i\,,
\end{equation}
so that the quantum Hamiltonian becomes effectively a function of the
operators $\hat q^i$ and~$\hat p_i$. On one-dimensional $\Sigma$ one
reobtains $\hat p^\sigma_i=\hat p_i$ as in the classical theory.

In the case where the fields $q^i$ form the coordinates of
a target space manifold $M$, one must take care of appropriate
integration measures in the state expansions, and of the fact that
$\delta(\bm{q}-\tilde{\bm{q}})$ is a density. The appropriate
Schr\"odinger representation would contain covariant, not partial,
differentiation operators on $M$.

\subsection{Hilbert space and probability interpretation}\label{Hilbert}

The $\Sigma$-dependent states $\left|\psi(\sigma^a)\right>$ are
elements of a Hilbert space $\mathcal{H}$. The essential algebraic
structure on a Hilbert space is a scalar product, i.e., a bilinear
form $\left<\cdot|\cdot\right>:\,\mathcal H\times\mathcal
H\rightarrow\mathbb R$. We may define such a scalar product in terms
of the wave functions corresponding to Hilbert-space states:
\begin{equation}
\left<\psi|\phi\right> = -i \int d\bm q \bar{\psi}\phi = -i \int d\bm q
\psi^{\dagger} \Gamma^0 \phi\,. 
\end{equation}
Note that the appearance of the gamma matrix $\Gamma^0$ of the local
Lorentzian tangent spaces guarantees that the scalar product maps to a
diffeomorphism scalar of $\Sigma$. The spinor indices of the
wave functions are suppressed, as are their $\Sigma$- and
field-dependence. 

The necessary requirement that the scalar product
should return a scalar function on $\Sigma$ results in its
indefiniteness: indeed,
\begin{equation}
\left<\psi|\psi\right> = -i \int d\bm q \psi^{\dagger} \Gamma^0 \psi \neq
\int d\bm q \psi^{\dagger} \psi\,,
\end{equation}
where the Dirac matrix mixes the spinorial components to prevent a
generically positive result. The indefiniteness is a feature of
quantization on manifolds $\Sigma$ of dimension $d>1$. For $d=1$, the
non-equality above becomes an equality (and $\psi^\dagger$ is simply
the complex conjugate). 

One of the consequences of this construction is the following:
the self-adjoint operators with respect to our scalar product are
no longer Hermitian with $\hat O^\dagger  = \hat O$. By definition,
self-adjoint operators satisfy
$\big<\hat O\psi\,|\,\phi\big>=\big<\psi\,|\,\hat O\phi\big>$ for all $\psi$ 
and $\phi$. Here this implies $\hat O$ is self-adjoint, if, and only
if,
\begin{equation}
\hat O^{\#} \equiv -\Gamma^0\hat O^\dagger \Gamma^0= \hat O\,. 
\end{equation}
But self-adjoint operators are not guaranteed to have real eigenvalues
because of the indefiniteness of the scalar product, which may produce
null states with $\left<\psi|\psi\right>=0$. Another condition is
needed: a self-adjoint operator is orthogonally diagonalizable 
with real eigenvalues if it
is a so-called Pesonen operator satisfying $\big<\psi\,|\,\hat
O\psi\big>\neq 0$ for all null states~$\psi$, see \cite{Bognar}. For
the special case $d=1$, we have $\Gamma^\sigma=i$ such that the relation
above selects Hermitian $\hat O$, as is the case in conventional quantum
mechanics.

The standard interpretation of quantum mechanics interprets the
squared modulus $|\psi|^2$ of the
Schr\"odinger wave function $\psi$ as a probability density. This is enabled
by the fact that the Schr\"odinger equation guarantees the constancy
of the total probability $\int d\bm x \psi^\dagger\psi$ in time. To
give a similar interpretation here, we need a similar
statement. Since the square $\left<\psi|\psi\right>$ is no longer
positive-definite for $d>1$ and hence no longer admits a probability
interpretation, we need to find another quantity that does. 
Consider the following vector current on $\Sigma$,
\begin{equation}\label{current}
j^a = - \int d\bm q \,\bar\psi\gamma^a\psi\,.
\end{equation}
We assume that our covariant Hamiltonian has essentially real
eigenvalues, meaning that it is self-adjoint with $\hat
H^{\#}=-\Gamma^0\hat H^\dagger\Gamma^0=\hat H$. This is consistent
because $\hat H$ is
directly related to~$\gamma^a\nabla_a$, and we have 
$\gamma^a{}^\dagger=\Gamma^0\gamma^a\Gamma^0$ and Hermitian
$i\nabla_a$. Then the Dirac form of relation (\ref{quevol}) is
sufficient to prove that $j^a$ is conserved,
\begin{equation}
\nabla_aj^a=0\,.
\end{equation}
Locally, a conserved current implies a conserved
charge: in normal coordinates on $\Sigma$,
\begin{equation}
\partial_{\sigma^0}\int d^{d-1}\sigma\,j^0 = 0\,.
\end{equation}

Thus we can interpret the integral of $j^0$ as the total probability
to find the fields in any configuration $\bm q$ anywhere on the
spatial part of $\Sigma$. This also means that
\begin{equation}
\rho(\sigma^a;\bm q) = -\bar\psi(\sigma^a;\bm
q)\gamma^0(\sigma)\psi(\sigma^a;\bm q)\,, 
\end{equation}
gives the probability density of finding the field configuration $\bm
q$ at a given point with coordinates $\sigma^a$ of $\Sigma$. In normal
coordinates, the probability density becomes $-\bar\psi\Gamma^0\psi =
\psi^\dagger\psi$, and is positive definite. In the
one-dimensional case $\Sigma\cong\mathbb{R}$, we can always find
global normal coordinates, so that we once again recover conventional
quantum mechanics.

The wave function $\psi(\sigma^a;\bm q)$ contains all the necessary
information to reconstruct the Schr\"odinger wave functional of the
conventional canonical approach in cases where the latter exists, as
has been shown in \cite{Kanatchikov:2000yh,Kanatchikov:2002yh}. This
suggests that the covariant canonical approach is at least as powerful
as the conventional one; more powerful, in fact, since it works on
backgrounds that do not allow the conventional spacetime
foliation by spatial hypersurfaces.

\subsection{Classical limit and Ehrenfest equations}\label{Ehrenfest}

To see how the classical limit emerges from our formalism, let us
consider the following Ehrenfest-type theorem, arising from the
expectation value of the Heisenberg equation of motion
(\ref{heisenevol}) above:
\begin{equation}
n(d)l^{-d+1}\nabla_a\big<\gamma^a\hat f\big>=\big<[\hat f,\hat H]\big>.
\end{equation}
Because the inner product is picture-independent, this equation in
particular holds in the Schr\"odinger picture. Assuming now a
classical covariant Hamiltonian of the form 
\begin{equation}
H=-\frac{\alpha}{2}p^a_i p^i_a+V(q^i)
\end{equation}
for constant $\alpha$, one finds $\hat H=\alpha \hat p^i\hat
p_i/(2dn(d)^2)+V(\hat q^i)$. We wish to evaluate the theorem for $\hat
f\mapsto \hat p^j$ as well as for $\hat f\mapsto \gamma_a \hat
q^j$. We use the commutator relations 
\begin{subequations}
\begin{eqnarray}
{[}\hat p_j,\hat H] & = & -idn(d)^2l^{-d+1}\frac{\partial V}{\partial q^j}(\hat q^i)\,,\\
{[}\gamma_a \hat q^j,\hat H] & = & i\alpha l^{-d+1}\gamma_a\hat p^j
\end{eqnarray}
\end{subequations}
to arrive at the following equations:
\begin{subequations}
\begin{eqnarray}
\left<\frac{\partial H}{\partial q^j}(\hat q^i,\hat p^b_k)\right> & = & -\nabla_a\big<\hat p^a_j\big>\,,\\
\left<\frac{\partial H}{\partial p^a_j}(\hat q^i,\hat p^b_k)\right> & = & \nabla_a \big<\hat q^j\big>\,.
\end{eqnarray}
\end{subequations}
Therefore, the classical covariant Hamiltonian field equations
(\ref{covHeqm}) are fulfilled within expectation values in the form of
Ehrenfest equations. Interestingly, no choice of the
normalizations of the length scale $l$ or of $n(d)$ was required to
obtain this classical result, showing once more the consistency of the
theory.


\section{Elementary applications}\label{sec.applications}

In this section we will give several simple applications of the covariant
canonical quantization method. This demonstrates the technique in some
detail, but, more importantly, yields a number of interesting results:
the Klein-Gordon equation arises as the wave equation of the
relativistic point particle without any need to refer to the 
relativistic energy-momentum relation; the quantization of any bosonic
field on an extended parameter space~$\Sigma$ with $d>1$ creates
spinorial wave functions; in particular, the Dirac equation emerges
from the Klein-Gordon Lagrangian, along with the prediction of a
fermion mass gap and a hierarchy of fermions that differ only by their
masses.

\subsection{Relativistic point particles -- the Klein-Gordon equation}

The simplest application of our quantization formalism is to the
relativistic mechanics of a point particle. Completely side-stepping
its usual derivation from the relativistic energy relation, we will
find that the quantum wave equation is the Klein-Gordon equation.

Consider the following action for fields $x^i:\mathbb{R}\rightarrow
M_{1,3}$ which describe the worldline embedding into a flat Minkowski
spacetime,
\begin{equation}
S=\int d\sigma \sqrt{-g_{\sigma\sigma}}\,\frac{1}{2}\left(m_1g^{\sigma\sigma}\nabla_\sigma x^i\nabla_\sigma x^j\eta_{ij}+m_2\right).
\end{equation}
Variation yields the equation of motion $\partial_\sigma
(\sqrt{-g_{\sigma\sigma}}g^{\sigma\sigma}\partial_\sigma x^i)=0$, and
also the gravitational constraint $g^{\sigma\sigma}\nabla_\sigma
x^i\nabla_\sigma x^j\eta_{ij}=m^2$. Both of these equations are needed
to show that  the above action is classically equivalent to the
standard action of the relativistic point particle, $m\int d\sigma
\sqrt{-\partial_\sigma x^i\partial_\sigma x^j\eta_{ij}}$, for two mass
parameters $m_1$ and $m_2$ which satisfy $m_1m_2=m^2$.

The covariant momenta are $p^\sigma_i=m_1g^{\sigma\sigma}\nabla_\sigma
x^j\eta_{ji}$, and yield the covariant Hamiltonian
${H=-\frac{1}{2m_1}g_{\sigma\sigma}p^\sigma_ip^\sigma_j\eta^{ij}-\frac{1}{2}m_2}$.
Using the Schr\"odinger representation of the momentum operators on
wave functions returns the quantum wave equation 
\begin{equation}
\left(-\frac{1}{2m_1}\square-\frac{m_2}{2}-i\partial_\sigma\right)\psi(\sigma ;x^i)=0\,,
\end{equation}
where the box denotes the d'Alembertian on $M_{1,3}$. We still have to
deal with the gravitational constraint on the classical momenta: in
its quantum version it reads $(\square+m^2)\psi=0$. This can be
satisfied consistently with the wave equation by choosing
$\partial_\sigma\psi(\sigma ;x^i)=0$. The resulting wave function is
then automatically independent of $\sigma$, and the equation for 
$\psi(x^i)$ becomes the Klein-Gordon equation on $M_{1,3}$.

The quantum wave equation is a Schr\"odinger equation with time
parameter $\sigma$, but independence of the wave function of this
parameter is forced by the constraint. In this sense, classical
reparametrization invariance directly implies the Klein-Gordon
equation.

\subsection{Free bosonic strings -- Weyl spinors}

One of the most characteristic features of covariant canonical
quantization is the fact that any bosonic field on a parameter space
of dimension $d>1$ produces spinor wave equations. To illustrate
this point in the simplest setup, we consider free bosonic strings
on a flat target space, given as maps $\Sigma\rightarrow M_{1,n}$
from the two-dimensional worldsheet $\Sigma$ into Minkowski 
space~$M_{1,n}$. We employ the Polyakov action 
\begin{equation}
S=-\int_\Sigma d^2\sigma\sqrt{-g} \, \frac{1}{2}g^{ab}\partial_a X^i\partial_b X^j \eta_{ij}\,.
\end{equation}
which is classically equivalent to the Nambu-Goto action that measures
the area of the string worldsheet, if the gravitational constraint
$g^{ab}g^{cd}\eta^*_{cd}=\eta^{*cd}$, where
$\eta^*_{ab}=\partial_aX^i\partial_bX^j\eta_{ij}$ is the pull-back of
$\eta$ to the worldsheet, is implemented.
The covariant momenta are derived as $p^a_i=-\partial^a X_i$ and
give rise to $H=-p^a_i p^i_a/2$. This leads to the Schr\"odinger
picture wave equation 
\begin{equation}
\left(l^{-1}n(2)\square + \gamma^a\partial_a\right)\psi(\sigma^b;X^\mu)=0\,,
\end{equation}
where $\square$ denotes the target space d'Alembert operator on
$M_{1,n}$.It is illustrative to use the separation ansatz
$\psi=\Psi(\sigma^a)\Phi(X^\mu)$ with spinorial $\Psi$ for the
wave function. This introduces a separation constant $M$, and
generates the two separate equations 
\begin{subequations}
\begin{eqnarray}
\left(\square + lMn(2)^{-1}\right)\Phi & = & 0\,,\\
\left(\gamma^a\partial_a-M\right)\Psi & = & 0\,. 
\end{eqnarray}
\end{subequations}
Thus the quantization of strings generates Weyl spinors on the
two-dimensional worldsheet, whose mass $M$ is linked to a Klein Gordon
equation on the target spacetime. 

As noted above, it is necessary to satisfy the gravitational
constraint. Classically, we may rewrite it as
$p^{ai}p^b_i-g^{ab}p^c_ip^i_c/2=0$ which is symmetric. Keeping the
symmetry we hence find the Schr\"odinger representation
\begin{equation}
\left(-\gamma^{(a}\gamma^{b)}+g^{ab}\right)\hat p^i\hat p_i=0\,.
\end{equation}
The expression in brackets is identically zero by the properties of
the Dirac algebra. Surprisingly the constraint is satisfied
automatically in the quantum theory. However, the quantization of the
string in this formalism has no discernible relation to string
theory, where the quantum requirement of the gravitational constraint
leads to the Virasoro algebra.

\subsection{Free scalar fields -- the Dirac equation}

The free scalar field $\phi:\Sigma_{1,3}\rightarrow\mathbb{R}$ on
a four-dimensional Lorentzian spacetime $\Sigma_{1,3}$ with metric
$g$ is governed by the Lagrangian
\begin{equation}
L=-\frac{1}{2}g^{ab}\partial_a\phi\partial_b\phi-\frac{1}{2}m^2\phi^2\,.
\end{equation}
The covariant momenta follow as $\pi^a=-\partial^a\phi$ and yield
$H=-\pi_a\pi^a/2+m^2\phi^2/2$. From quantization we hence obtain the
Schr\"odinger picture wave equation 
\begin{equation}
\left(-\frac{2n(4)}{l^3}\partial_\phi^2 + \frac{l^3 m^2}{2n(4)}\phi^2 - \gamma^a\partial_a\right)\psi(\sigma^a;\phi) = 0\,. 
\end{equation}
To illustrate the consequences of our theoretical construction, it is
useful to consider a separation ansatz $\psi=\Psi(\sigma^a)\Phi(\phi)$
for the wave function, where only $\Psi$ is taken spinorial. This
introduces a separation constant $M$, and the wave equation generates
two separate equations of the form
\begin{subequations}\label{sceq} 
\begin{eqnarray}
\left(-\frac{2n(4)}{l^3}\partial_\phi^2 + \frac{l^3 m^2}{2n(4)}\phi^2\right)\Phi & = & M\Phi\,,\label{harmosci}\\
\left(\gamma^a\partial_a-M\right)\Psi & = & 0\,. \label{emergdirac}
\end{eqnarray}  
\end{subequations}
Thus the spacetime dependence of the wave function is described by a
Dirac equation with mass $M$. However, this mass is not unconstrained:
a spectrum of allowed masses is generated by the first equation, which
is just the Schr\"odinger equation of a one-dimensional harmonic
oscillator with frequency $\omega=2m$. The
mass spectrum is therefore given by
\begin{equation}
M_k = m \left(2 k+1\right)
\end{equation}  
with integer $k\geq 0$.
While this spectrum does not look particularly appealing
phenomenologically, it should be noted that we obtained the prediction
of a mass hierarchy of otherwise identical Dirac particles from the
non-interacting Klein-Gordon Lagrangian only. With the addition of
interactions to the scalar action, more complex mass spectra could be
generated, leading to the possibility of obtaining the generations of
the Standard Model from a suitably tuned interacting \emph{scalar}
Lagrangian. Another important prediction from equations~(\ref{sceq})
is the existence of a mass gap for the fermions: $M=0$ is generally
not a solution if the harmonic potential $V(\phi)\sim\phi^2$ is
replaced by a generic potential $V(\phi)$.

\subsection{Local gauge invariance -- gauge fields}

We have seen that the quantization of pure scalar field models
generates fermionic particles with an allowed mass spectrum given by
the covariant Hamiltonian of the scalar field. Phenomenological
relevance additionally requires these fermions to be charged. A
mechanism to serve this purpose has been identified by Weyl
\cite{Weyl:1929} a long time ago, and we will now demonstrate its effect.

The basic observation underlying this mechanism is the invariance
of the interpretationally relevant probability current
$j^a=-i\left<\psi\,|\,\gamma^a\,|\,\psi\right>$ defined in
(\ref{current}) under local phase shifts of the wave function,
so-called gauge transformations. These are transformations
$\psi\mapsto e^{ie\Lambda}\psi$ under a function
$\Lambda:\Sigma\rightarrow\mathbb{R}$. The quantum evolution
equation (\ref{quevol}), however, is only invariant under global
gauge transformations with constant $\Lambda$. If, in addition,
local invariance is required, we have to amend this equation by the
introduction of a gauge field: 
\begin{equation}
\left(\hat H - n(d)l^{-d+1}(\sla\nabla-ie\sla A)\right)\left|\psi\right>=0\,.
\end{equation}
It now follows that if $\left|\psi\right>$ is a solution of this
equation, then so is the locally gauge transformed
$e^{ie\Lambda}\left|\psi\right>$, as long as the gauge field
transforms at the same time as $A_a\mapsto A_a+\partial_a\Lambda$. The
resulting equation is essentially the Dirac equation for particles of
mass $\big<\hat H\big>$ and charge~$e$.

To make this statement more precise, observe that in order to
speak about the spectrum of the theory we do not
require knowledge about the probabilities for the original scalar
fields, but only about the generated fermions. This means we may
consider the integrated expectation value of the quantum evolution
equation as a functional of $\psi$, 
\begin{equation}
S[\psi]=\int_\Sigma d^d\sigma\left<\psi\right|\left(-n(d)^{-1}l^{d-1}\hat H + \sla\nabla-ie\sla A\right)\left|\psi\right>
\end{equation}
in which the wave function's dependence on the original scalar fields
is effectively integrated out. Indeed, if we assume the wave equation
has been solved by a product ansatz ${\psi=\Psi(\sigma^a)\Phi(\phi)}$
as in the previous examples, we find  
\begin{equation}
S_\Phi[\Psi]=\left(\int d\phi \Phi^\dagger\Phi\right) \int_\Sigma d^d\sigma\,\bar\Psi\left(\sla\nabla-ie\sla A-\tilde M\right)\Psi\,,
\end{equation}
for one of the rescaled mass eigenvalues $\tilde
M=n(d)^{-1}l^{d-1}\int d\phi \Phi^\dagger\hat H\Phi/\int d\phi
\Phi^\dagger\Phi$ in the mass spectrum generated by the scalars'
covariant Hamiltonian. So integrating out the original
scalar field freedom, because it is not observable, returns precisely
the Lagrangian theory for the Dirac field with mass $\tilde M$.

Now consider a multiplet of $N$ scalar fields $\phi_i$ with
  a Lagrangian which is invariant under the action of some nonabelian
  subgroup $G$ of $SO(N)$. In this case, the covariant Hamiltonian
  will inherit the $G$-invariance, resulting in a
  degeneracy in the spectrum of the theory. The energy levels $M_n$
  can then be labelled by the irreducible representations of $G$, with
  the degeneracy of each level given by the dimension $d_n$ of the
  irreducible representation under which it transforms. A solution of
  the quantum evolution equation can then be decomposed in terms of
  eigenfunctions of the covariant Hamiltonian as
\begin{equation}
\psi(\sigma,{\bm\phi}) = \sum_{n=0}^\infty \sum_{\alpha=1}^{d_n}
\Psi_{n,\alpha}(\sigma)\Phi_{n,\alpha}({\bm\phi})
\end{equation}
  Putting this decomposition into the definition of $S[\psi]$ above
  and using the orthogonality relation
\begin{equation}
\int_\mathcal{M} d{\bm\phi} \Phi^\dag_{n,\alpha}({\bm\phi})
\Phi_{m,\beta}({\bm\phi}) = \delta_{m,n}\delta_{\alpha,\beta}
\end{equation}
  for the wavefunctions, we arrive at the action for $\Psi$
\begin{equation}
S[\Psi] = \sum_{n=0}^\infty \int_\Sigma d\sigma \bar{\Psi}_{n,\alpha}
\left(\sla\nabla -\tilde M_n\right)\Psi_n^\alpha
\end{equation}
where each kind of fermion has a mass $M_n$ and is invariant under a
SU($d_n$) symmetry in addition to the U(1) symmetry discussed
above. Using the same arguments as before, these SU($d_n$)
symmetries should also be gauged, giving rise to a spectrum of
nonabelian gauge symmetries.

In the case of a finite scalar symmetry group $G$, there
is only a finite number of irreducible representations, and
correspondingly the gauge group of the fermionic theory then is a
finite product of SU($N$) gauge groups. For continuous $G$, an
infinite product of SU($N$) factors ensues. Groups that could
give a Standard Model-like gauge group SU(3)$\times$SU(2)$\times$U(1) 
(although we haste to point out that this simple model does not give
the chiral couplings of the Standard Model) include the point groups
$T_d$ and $O$.

Hence, covariant canonical quantization with the additional
requirement of local gauge invariance is able to produce, from a
scalar field Lagrangian, all particles (so far) observed in Nature,
namely fermions and gauge fields. While their masses are constrained
by the covariant Hamiltonian of the original classical scalar field
theory, the charges are additional input at the quantum level of the
theory. In order to obtain a more complete theory, appropriate
gauge-invariant dynamics for the gauge fields have to be added at
this stage, which is a freedom that we have.


\section{Discussion}\label{sec.discussion}

Considering fields as maps from a parameter space $\Sigma$ to a
target space $M$, we have constructed a covariant quantization method
that keeps the diffeomorphism invariance between the parameters of
$\Sigma$ intact. Covariant canonical quantization is based on the
classical Hamiltonian theory developed  by de
Donder~\cite{deDonder:1935} and Weyl~\cite{Weyl:1935} which makes use
of a finite-dimensional multi-symplectic phase space, where every
field has a set of conjugate momenta associated to each of its partial
derivatives. The classical theory is completely equivalent (in
the sense of generating the same solutions) to the conventional
Hamiltonian point of view, avoiding, however, the explicit
breaking of diffeomorphism invariance that arises from 
singling out a time coordinate normal to an assumed foliation of
$\Sigma$ by spatial hypersurfaces. 

We have introduced the notion of a covariant Poisson bracket within
the classical theory, before applying two well-motivated quantization
postulates. The first postulate replaces, according to Dirac's
argument, the Poisson bracket of phase space functions by the
commutator of corresponding operators acting on some Hilbert
space. The second postulate is geometrically motivated: on parameter
spacetimes $\Sigma$ of dimension $d>1$ it introduces the Clifford
algebra of Dirac matrices into the quantum theory. The construction is
such that covariant canonical quantization coincides with conventional
canonical quantization in $d=1$, i.e., when the fields depend only on
time.

The two quantization postulates, applied to the classical de
Donder--Weyl theory, for the first time allow the derivation of a
quantum evolution equation in terms of the covariant Hamiltonian,
which had been conjectured before on the mere basis of
analogies~\cite{Kanatchikov:1996fx,Navarro:1998en}. This evolution
equation effectively unifies the Dirac and the Schr\"odinger
equations. We have further developed the theory in the covariant
Schr\"odinger picture, including a discussion of the representation of
the field and multi-momentum operators, and the relevant Hilbert
space. Diffeomorphism invariance requires an indefinite inner product
on the Hilbert space whose consequences for the diagonalizability of
operators and their eigenvalues have been discussed. We have also
provided a probability interpretation for the fields' wave function. 

Further development of the theory could progress in several
directions. One of the obvious questions concerns the
Heisenberg picture for the theory. The formal solution for the
evolution operator follows from equation~(\ref{U1}) as the
path-ordered exponential
$U(\sigma,\sigma_0)\sim\mathcal{P}\exp\left(l^{d-1}/(dn(d))\,\int_{\sigma_0}^\sigma
  H \gamma_a d\xi^a\right)$. As such it is only locally defined, and
path-dependent, at least on generic curved spacetimes $\Sigma$ where
we had to specify the path along which the exponent is integrated
by using geodesic normal coordinates on sufficiently
small neighborhoods. This raises questions
about the possibility of developing scattering theory in these
cases. Another question is that 
of the quantization of gravity. Although we have made use of
gravitational constraints in two of the examples, it is not so clear
how the theoretical setup could be consistent with additional dynamics
for the background metric on the parameter space $\Sigma$. 

We have discussed a number of elementary applications of the formalism
with some surprising and very interesting results. The quantization of
the relativistic point particle immediately yields as quantum wave
equation the Klein-Gordon equation, without conceptually employing the
relativistic energy momentum relation. Though
there is no readily discernible connection to string theory, it
is interesting 
to note that the quantization of bosonic strings produces Weyl spinors
on the worldsheet whose mass is linked to a Klein-Gordon equation on
the target space. This fact expresses one of the most characteristic
features of covariant canonical quantization, namely that the
quantization of any field on $\Sigma$ with $d>1$ produces a spinorial
wave equation. 

We have shown that a purely scalar classical Lagrangian can produce,
upon quantization, a theory of Dirac fermions interacting with gauge
fields. The latter come into the theory by requiring the local gauge
invariance of the interpretationally relevant probability current also
on the level of the wave equation. This means that the basic equations
of the quantum field theory of the Standard Model may emerge from a
classical scalar field theory as wave equations. Several
interesting results are obtained along with this mechanism: 

The emergence of spinor fields as a purely quantum phenomenon bypasses
the usual need for a semi-classical treatment of the Dirac equation,
which emerges as a fully quantum equation from the very
beginning. More intriguingly, on the phenomenological side, the
procedure of covariant canonical quantization provides a new mechanism 
to unify particles of different masses, which leads to an (at least 
qualitative) explanation of the 
generations of the Standard Model in terms of a self-interacting
scalar field. It should be noted that in this framework fermionic
masses are generated without a Higgs mechanism. Instead, the mass and
self-interaction of the underlying scalar field manifest themselves by
generating a mass spectrum for the effective fermion field which is
the spacetime part of the scalar field's quantum wave function. 

It is interesting to speculate where a scalar model, complicated
enough to yield the standard model upon quantization, could come from
in the first place. One possibility seems to be the compactification
of a higher-dimensional bosonic, maybe gravitational, theory. Such
compactifications generally produce a large number of scalar fields as
shape moduli of the internal manifold, with self-interactions through
some effective potential. If such a potential had a minimum, one would
expect it at some negative value, due to a geometric no-go
theorem~\cite{Gibbons:1984kp} (see also the discussion
in~\cite{Townsend:2003fx}), thus generating a discrete mass spectrum
for the fermionic fields in the quantum theory which could have
phenomenological relevance. More speculatively still, the fact
that the quantization of a purely scalar classical theory necessarily
leads to a fermionic quantum theory might be indicating some sort of
semi-classical supersymmetry at work behind the scenes. 

Covariant canonical quantization clearly does not replace
conventional quantum field theory; rather it adds a first
quantization to the usual procedure, which then literally becomes a
second quantization: first, a scalar model generates spinorial
quantum wave equations which, after integrating out the unobservable
scalar degrees of freedom, become the classical equations of motion
underlying the Standard Model. Then the quantization of these
equations proceeds in standard quantum field theoretical fashion. The
intriguing result of this investigation is the fact that the Standard
Model's classical equations of motion for fermions and gauge fields,
along with the prediction of discrete mass spectra of identically
charged particles, are generated from the quantization of a purely
scalar classical field theory. This gives reason to hope that
covariant canonical quantization might find further uses and
applications.


\acknowledgments

MNRW thanks Klaus Fredenhagen and Stefan Hofmann for very useful
discussions. He acknowledges financial support from the German
Research Foundation (DFG), the German Academic Exchange Service
(DAAD), and the European RTN program MRTN-CT-2004-503369.
GMvH acknowledges partial financial support from the Natural
Sciences and Engineering Research Council of Canada and the
Government of Saskatchewan.


\end{document}